\documentclass[twocolumn]{article}
\usepackage[top=2cm, bottom=2.8cm, left=2cm, right=2cm]{geometry}

\usepackage{graphicx}
\usepackage{mathptmx} 
\usepackage{mathtools}
\usepackage{hyperref}
\usepackage[font=scriptsize]{subfig}
\usepackage[font=scriptsize,labelfont=bf]{caption}
\usepackage{xcolor}
\usepackage{amssymb}
\usepackage{amsmath}
\usepackage{tabularx}

\usepackage[super,numbers]{natbib}
\usepackage{sans}
\usepackage{authblk}
\providecommand{\keywords}[1]{\textbf{\textit{keywords---}} #1}
\begin{document}

\title{The effect of varying degrees of stenosis on transition to turbulence
in oscillatory flows}

\markright{Effect of stenosis degree on oscillatory flow transition}

\author[1]{Kartik Jain
  \thanks{Corresponding Author; E-mail: \texttt{k.jain@utwente.nl}}}

\affil[1]{
  \small Faculty of Engineering Technology,
  University of Twente, P.O. Box 217, 7500AE
  Enschede, \textsc{The Netherlands} 
}

\date{}

\maketitle

\begin{abstract}
  Many complications in physiology are associated with a deviation in flow in
  arteries due to a stenosis.
  The presence of stenosis may transition the flow to weak turbulence. 
  The degree of stenosis as well as its configuration whether symmetric or
  non-symmetric to the parent artery influences whether the flow would stay
  laminar or transition to turbulence.
  Plenty of research efforts focus on investigating the role of varying degrees
  of stenosis in the onset of turbulence under steady and pulsatile flow
  conditions.
  None of the studies, however, have focused on investigating this under
  oscillatory flow conditions as flow reversal is a major occurrence in a
  number of physiologic flows, and is of particular relevance in cerebrospinal
  fluid (CSF) flow research.
  Following up on the previous work in which a $75\%$ stenosis was studied,
  this contribution is a detailed investigation of the role of degrees of
  stenosis on transition in an oscillatory flow.
  A cylindrical pipe with $25\%$, $50\%$ and $60\%$ reductions in area in
  axisymmetric and eccentric configurations is studied for transition with $3$
  different pulsation frequencies of a purely oscillatory flow.
  Cycle averaged Reynolds numbers between $1800$ and $2100$ in steps of $100$
  are studied for each configuration resulting in $72$ simulations each
  conducted on $76\,800$ CPU cores of a modern supercomputer.
  It is found that a higher degree of stenosis and eccentricity causes earlier
  transition to turbulence in oscillatory flow.
  The results further demonstrate that a higher frequency of oscillation
  results in larger hydrodynamic instability in the flow, which is more
  prominent in smaller degrees of stenosis\footnote{\textbf{THIS ARTICLE HAS BEEN ACCEPTED FOR PUBLICATION IN BIOMECHANICS AND MODELING IN MECHANOBIOLOGY on 25-Mar-2022} }.

\keywords{Lattice Boltzmann Method, transitional flow, turbulence, stenosis, hydrodynamic instability}
\end{abstract}

\section{Introduction}

A stenosis refers to obstruction in a blood vessel or an anatomical segment
like the subarachnoid space, spinal canal or respiratory airway.
A stenosis can induce turbulence and flow separation in the blood vessel
or the anatomical segment as a result of adverse pressure gradients.
The presence of turbulence can eventually lead to secondary complications
to the physiology in question.
The onset of turbulence itself is likely to be influenced by the degree of
stenosis itself as well as the nature of the flow, whether steady, pulsatile or
oscillatory.
For example, stenoses in the carotid arteries are well known risk factor for
ischemic stroke~\citep{fairhead05a}.
The degree of stenosis has been established as a factor for stroke risk and
indication for intervention~\citep{qureshi07a}.
A recent review by~\citet{brinjikji16a} delineates the current advances in
plaque imaging and association between degree of stenosis and the progression
of plaque, while~\citet{berger00a} reviewed studies of stenotic flows with a
focus on atherosclerosis.

Previous studies have extensively investigated the complexity of fluid dynamics
in stenosed vessels both numerically~\citep{varghese1, varghese2, samuelsson}
and experimentally~\citep{gidden1, gidden2}.
While studies of steady stenotic flow provide a general guidance about the
nature of flow in such vessels, researchers try to assess the parts of cardiac
cycles with most turbulent nature of the flow.
In particular in the late systole when the flow decelerates, the flow has been
shown to be highly disturbed, which stabilizes during the early diastole or
acceleration phase of the cardiac cycle.
The early studies of~\citet{yoganathan86a, bluestein95a} investigated in detail
as to what degree of stenosis causes the flow to become turbulent.

{\color{black}
  Most of the studies of stenotic flows usually focus on uni-directional flows.
  This is natural because the areas where a stenosis are most relevant have a
  unidirectional steady or pulsatile flow.
  Earlier experiments of~\citet{gidden1} investigated post stenotic flow in a
  tube using LDA at mean Reynolds number of 600 and Womersley number $W_o = 7.5$.
  They found strong turbulence and flow separation downstream of the $75\%$
  stenosis.
  \citet{ding21a} recently studied pulsatile flow in a 2D stenosed pipe using PIV
  experiments with a mean Reynolds number of 1750 and Womersley number of $W_o =
  6.15$.
  They employed $25\%$, $50\%$ and $75\%$ degree of stenosis and found increasing
  turbulent intensities with increasing constrictions.
  \citet{xu17a} carried out experiments in a sinusoidal shaped pulsatile pipe
  flow and observed that the critical Reynolds number increased with an increase
  in the Womersley number until $W_o= 2.5$. 
  Then the critical Reynolds number reduced with a further increase in the
  Womersley number until it reached to $W_o=12$.
}

Some parts in the circulation however do experience a flow reversal.
For example, the waveform of blood flow through arteries has been shown to
change abruptly when passing from the thoracic aorta into the abdominal aorta
in humans, exhibiting regions of flow reversal at the end of systole.
\citet{holenstein88a} developed a lumped parameter model of such a flow and
referred to it as \emph{triphasic} flow with positive-negative-positive parts
of the cardiac cycle.
Another form of such a triphasic flow is a purely oscillatory flow, known to
occur mainly in the central nervous system (CNS) namely the circulation of the
cerebrospinal fluid (CSF).
Another purely oscillatory flow is the airflow in the respiration system.
In a purely oscillatory flow the underlying factors that trigger turbulence,
the distribution of turbulence in various parts of the cycle, as well as the
stabilization mechanisms are different than those from a steady or pulsatile
flow with no reversal component.
Earlier experiments by~\citet{hino76a, hino83a} have investigated detailed
characteristics of turbulence in oscillatory pipe flow.
Other similar studies~\citep{sarpkaya, yellin, ohmi2} addressed such aspects
but also in simple pipes without any distortion that would resemble an
anatomical geometry.

In a recent work~\citep{jain20a} we investigated transition to turbulence in a
purely oscillatory flow with three pulsation frequencies in a pipe with $75\%$
area reduction in eccentric and axisymmetric configurations.
While the role of symmetry of the geometry and the pulsation frequencies in the
onset of turbulence was the focus of that study, the degree of stenosis itself
was not investigated.
The present contribution investigates the characteristics of turbulence in a
stenosed pipe with area reductions of $60\%$, $50\%$ and $25\%$.
All three degrees of stenosed pipes are studied in two configurations: one that
is symmetric to the principal axis of the pipe while the other which is offset
by $0.05$ diameters of the pipe to introduce an eccentricity or a geometric
aberration.
Three different flow oscillation frequencies with Womersley numbers ranging
between $\sim 5-8$ are studied in all the degree and configurations of the
stenosis to explore detailed characteristics of flow transition.
Reynolds numbers between $1800$ and $2100$ are investigated in steps of $100$
to compare the characteristics of flow transition with varying degrees of
stenosis against the previous study~\citep{jain20a}.

{\color{black}
  The three main focal points of the study namely \emph{degree of stenosis},
  \emph{pulsation frequencies} and \emph{symmetry/asymmetry of the geometry}
  are representative of a broad range of CSF and respiratory flows.
  For example stenosis of the lumbar spine ranges from various degrees within
  the range studied here~\citep{genevay10a} while the obstruction to CSF caused
  by Chiari I malformation also lies in this range~\citep{linninger16a}.
  Subglottic stenosis is a narrowing of a specific portion of the windpipe
  (trachea) known as the subglottis (just below the vocal cords).
  These are classified as grade 1 stenosis with a luminal obstruction $<50\%$,
  grade 2 and 3 stenosis with $51-70\%$ and $71-99\%$ obstruction
  respectively~\citep{jefferson20a}.
  The pulsation frequencies range from a Womersley number of approximately 5
  (CSF) to 11 (respiratory airflow), which corresponds to the pulsation
  frequencies studied in this work.
  Occurrence of stenosis in a symmetric fashion is rare due to the complexity
  of the anatomy and most obstructions are found in a highly asymmetric
  fashion.
  Similarly, the Reynolds numbers, mainly due to the presence of stenosis can
  reach up to $2000$ in CSF and $1500-8000$ in respiratory airways.
  The present study is thus a generalization of stenotic flow in these two
  applications.
  The idea is to introduce representative anatomic and physiologic complexity
  while maintaining simplicity that could allow prospects for future studies in
    identifying the chances of turbulence.
  The goal thus is that these canonical studies serve as benchmark for future
  comparisons and thus a patient specific anatomic case has not been studied.
}

For the absence of a clear definition for a flow regime that is neither
laminar nor turbulent, we have addressed the flow in this study
as transitional.
Thus, if fluctuations occur in localized parts of the domain during parts
of the cycle, we term the flow transitional.
This is in accordance with several such studies in this
direction~\citep{varghese2, samuelsson, yellin, jain20a, jain20b}.
The onset of a fully developed turbulence is not pursued and the focus is on
exploring the critical Re at which the flow leaves a laminar regime for a
particular degree of stenosis, and a specific pulsation frequency (Womersley
number).
The Reynolds number at which the flow transitions is judged on the basis of
hydrodynamic instabilities in the velocity field and deflection of the jet
downstream of the stenosis.

{\color{black}
  We have employed the lattice Boltzmann method (LBM) for the DNS of this study
  as the method scales well on parallel computers {\textendash} allowing for
  accurate capture of transitional flow characteristics.
  The LBM when setup properly~\citep{jain20b} is a very appropriate second
  order accurate method for the DNS of flow.
  While the order of accuracy is lower than competitive higher order methods,
  the strict control of numerical viscosity in LBM allows for a relatively
  easier capture of flow fluctuations.
  A comparison of LBM with other second-order accurate methods~\citep{marie}
  has found that the numerical dissipation in LBM, even at the scales of grid
  spacing and the numerical dispersive effects are relatively smaller.
}
\section{Methods}
  \begin{figure}%
    \centering%
    \subfloat[Bisecting plane showing different degrees of stenosis with and without eccentricity]{
      \includegraphics[width=0.95\columnwidth]{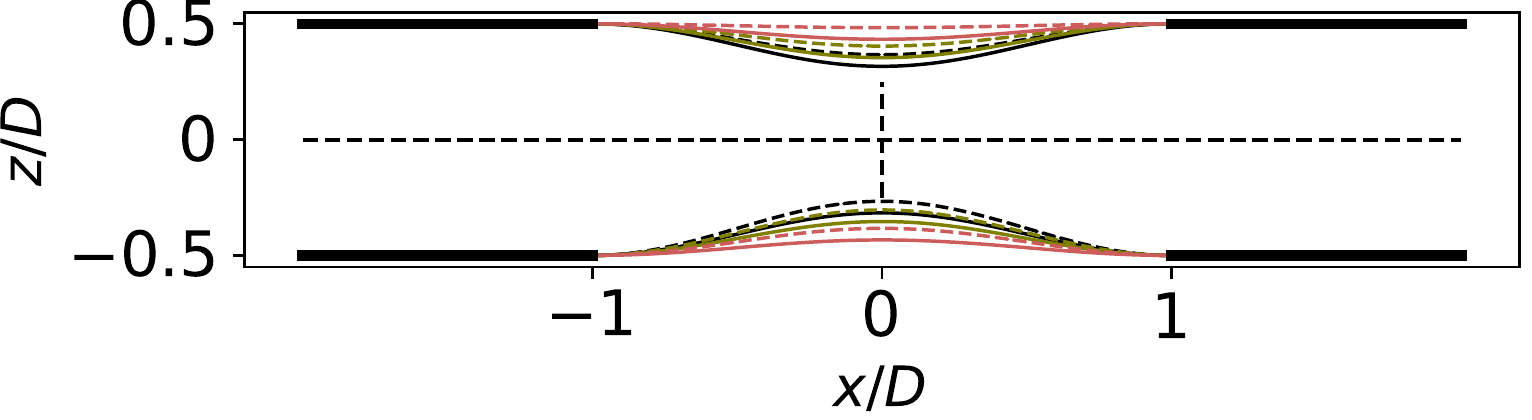} \label{fig:stenecc}
    }
    \hspace{10pt}
    \subfloat[Cross sectional view ]{
      \includegraphics[width=0.3\columnwidth]{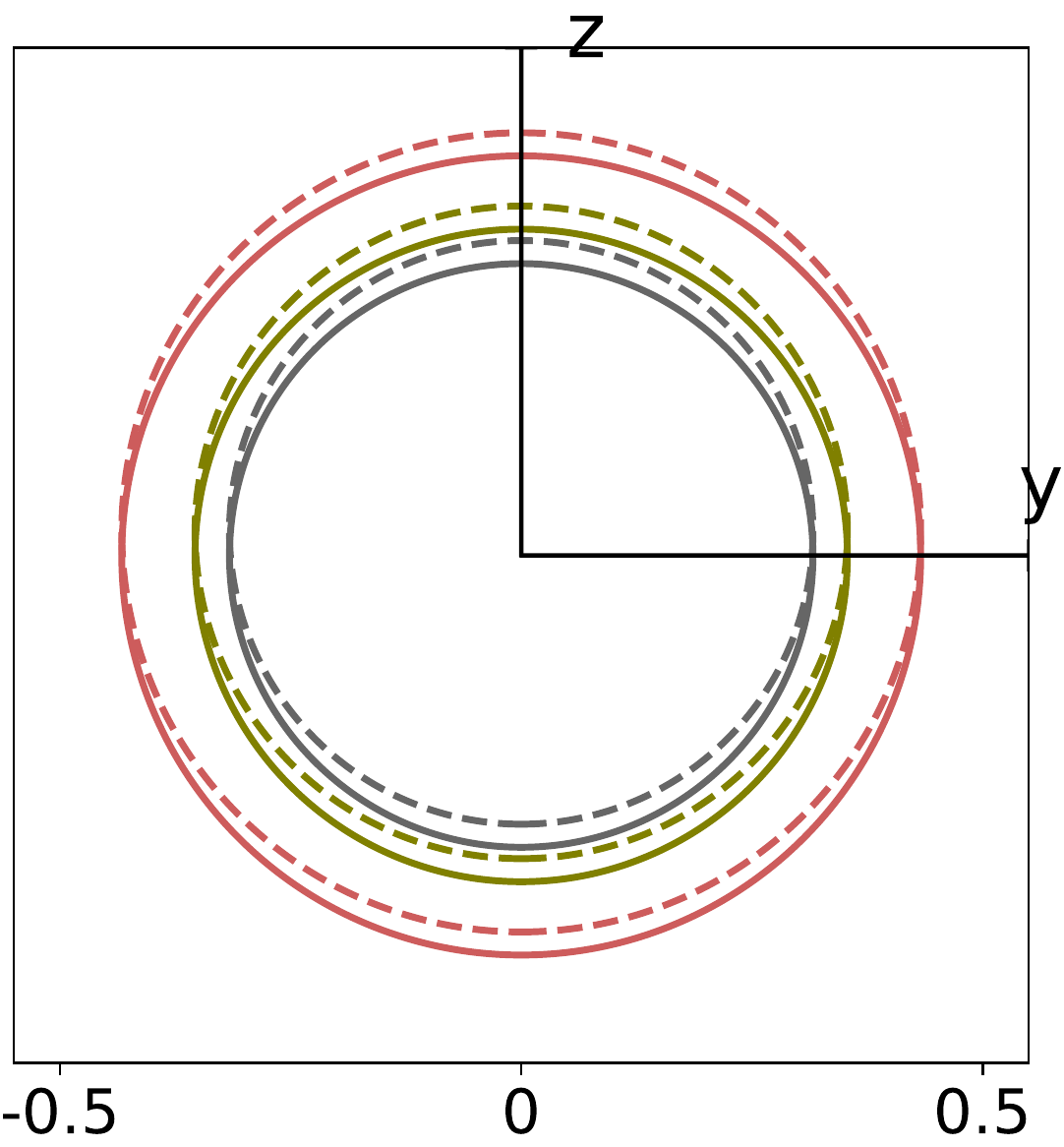} \label{fig:circle}
    }
    \caption{
      Top: A bisecting plane showing the principal pipe with varying degrees of
      stenoses where gray, green and red respectively represent $60\%$, $50\%$
      and $25\%$ reduction by area.
      Corresponding dotted lines show the offset of $0.05$ diameters of the
      principal pipe to introduce an eccentricity in each case.
      Bottom: Cross sectional view of different degrees of stenoses.
    }
    \label{fig:model}
  \end{figure}
  \begin{figure}%
    \centering%
      \includegraphics[width=\columnwidth]{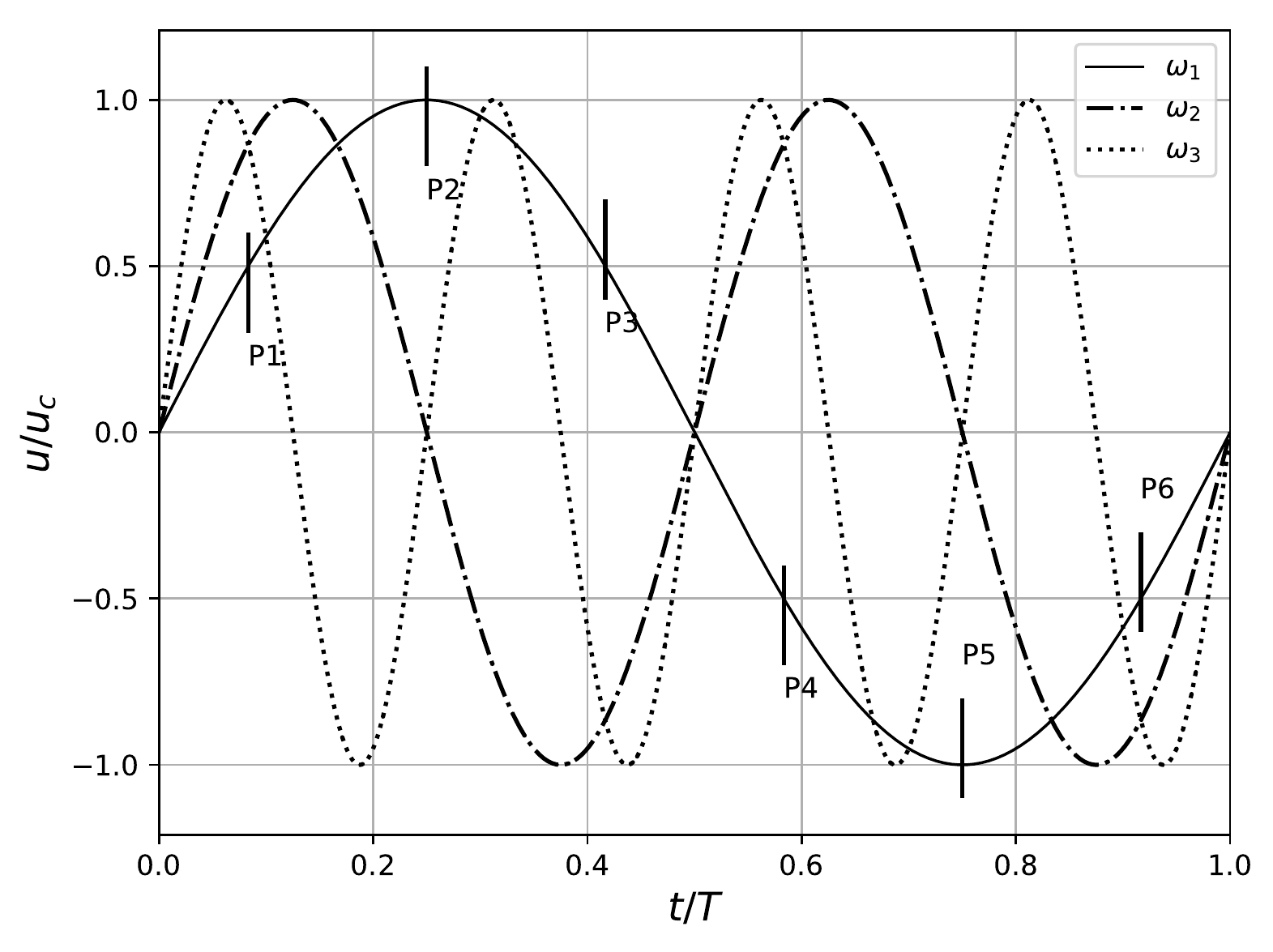}         
    \caption{Axial centerline velocity at the vessel inlet with three different
    oscillation frequencies, $\omega_1$, $\omega_2$ and $\omega_1$.
    Observations were made during $6$ equidistant points along the sinusoidal
    cycle (shown only for $\omega_1$)}
    \label{fig:sinus}
  \end{figure}
  The geometry of the stenosis was adopted from the experiments
  of~\citet{gidden2} (shown in figure~\ref{fig:model}).
  Stenosis of varying degrees in axisymmetric and eccentric configurations
  were created using the Blender software previously by~\citet{jana_bs}.
  The gray, green and red lines respectively demonstrate $60\%$, $50\%$ and
  $25\%$ reduction in area whereas there dotted counterparts indicate the
  eccentricity that was introduced.

  Similar to previous study~\citep{jain20a} the pre and post-stenotic regions
  of the vessel were \emph{equally} extended by $12$ vessel diameters to
  account for the reversing nature of the flow, which resulted in a total pipe
  length of approximately $26$ diameters.
  Due to flow reversal, the fluid travels equally on both sides of the stenosis
  throat, and the location of stenosis in the pipe is thus expected to
  influence the onset of transition as well as propagation of formed turbulent
  bursts.

  \subsection*{Oscillatory flow}
  The oscillatory nature of the flow can be described by the Womersley solution
  with a negative (flow reversal) component.
  The Womersley solution for laminar, pulsatile flow through rigid tubes
  is defined as:
    \begin{equation}              \label{eq:womer}
      \begin{split}           
        \begin{drcases}
            \frac{u_x}{u_c} & = A \Bigg[1 - \frac{J_{0}(i^{3/2} \alpha
            2r/D)}{J_0(i^{3/2} \alpha )} \Bigg] sin(\omega t),              \\
            \frac{u_y}{u_c} & = 0,                                          \\
            \frac{u_z}{u_c} & = 0
        \end{drcases}
      \end{split}
    \end{equation}
  where $u_c$ is the half-cycle-averaged inlet centerline velocity, A and
  $\omega$ respectively are the amplitude and angular frequency of pulsation,
  $J_0$ is the Bessel function of first kind and zeroth order, and $\alpha$ is
  the non-dimensional Womersley parameter ($= \frac{1}{2}D \sqrt{\omega/\nu}$,
  where $\nu$ is the kinematic viscosity).
  The Womersley parameter defines the extent to which the laminar profile
  departs from quasi-steadiness. 
  This effect becomes significant when the Womersley parameter, $\alpha = 3$.
  Three different frequencies of pulsation were studied where 
  $\omega_2=2 \omega_1$ and $\omega_3=4 \omega_1$.
  The forward and backward flow are referred to as \emph{blowing} and
  \emph{suction} stages respectively in the following.

  Figure~\ref{fig:sinus} depicts the flow waveform and the 6 time intervals
  of T/6 where flow quantities were analyzed.
  Like the studies of~\citet{jain20a}, the flow profile here consists of a
  negative component which accounts for the flow reversal, and the mean flow in
  this case is thus zero describing a purely oscillatory flow.
  The Reynolds number was based on the main vessel diameter, D, and the mean
  inlet centerline velocity of \emph{half the cycle}, $u_c$.
  The Re was varied from $600$ to $2100$ in steps of $\delta Re = 100$ for both
  the configurations of the stenosis and three pulsation frequencies.  
  This resulted in a total $90$ sets of simulations.
  %
  {\color{black}
    The value of A and $\alpha$ in equation~\ref{eq:womer} were $0.667$ and
    $5.4$ respectively.
    For the lowest pulsation frequency $\omega_1$ the Womersley parameter
    $\alpha_1$ was chosen as $5.4$ keeping the frequency of oscillations of the
    cerebrospinal fluid (CSF) in view.
    To attain the pulsation frequencies of $\omega_2=2 \omega_1$ and
    $\omega_3=4 \omega_1$, the corresponding Womersley parameters became
    $\alpha_2=7.636$ and $\alpha_3=10.8$ respectively.
    Zero pressure and velocity were set at every lattice cell as initial
    conditions.
    A zero pressure gradient was maintained at the outlet of the pipe which
    translates to an extrapolation of incoming populations at the
    outlet~\citep{junk2011}.
    The flow waveform shown in figure~\ref{fig:sinus} was prescribed at the
    inlet.
    Flow was allowed to develop for two initial cycles before analysis was
    performed.
  }

  \subsection*{Direct Numerical Simulation}
  {\color{black}
  The adaptable poly engineering simulator (APES) framework~\citep{apes, parco,
  parco2016} was chosen for the simulations which contains a full tool-chain of
  simulation software ranging from the mesh generator
  \emph{Seeder}~\citep{seeder} and the LBM solver \emph{Musubi}~\citep{musubi}.
  Modules within APES are available as an open source software tool for
  academic research \footnote{https://apes.osdn.io}.
  The \emph{Musubi} LBM solver is mainly managed at the Institute of Software
  Methods for Product Virtualization within the German Aerospace Center (DLR)
  in Dresden, \textsc{Germany}.
  Extensive development and use of the solver for applications in physiology is
  done at the University of Twente, \textsc{The Netherlands}.

  Extensive verification of the LBM solver \emph{Musubi} has been performed 
  in previous works~\citep{musubi}.
  About 30 verification test cases are executed automatically every night that
  ensure that the physics computation and parallel as well as serial performance
  of the solver compares well with the benchmarks.
  Several efforts in the direction of validation have been placed, see for
  example a comparison against experiments reported in~\citet{johannink}.
  A comprehensive validation for the simulation of physiologic flows in
  transitional and turbulent regime was conducted using the benchmark set out
  by the US Food and Drug Administration (FDA)~\citep{jain20b}.
  For each new biomedical application, comparisons against \emph{in vitro} and
  \emph{in vivo} measurements are performed during the continuous use and
  development of \emph{Musubi}.
  See for example, a recent work conducted in parallel in which simulation
  results were compared against \emph{in vitro} experiments of nasal
  airflows~\citep{hebbink22a}.
  }

  Due to the simplicity of the geometry studied in this work, the employment of
  the Bhatnagar-Gross-Krook (BGK) relaxation scheme was sufficient.
  The BGK relaxation parameter was set to $\Omega = 1.84$ in the present study
  that keeps the lattice Mach number within the stability limits of the
  LBM~\citep{junk}.
  Further details on these aspects can be referred elsewhere~\citep{jain20b}.

  The spatial and temporal resolutions were chosen as
   $\delta x = 32 \times 10^{-4}$ and $\delta t = 7.5 \times
  10^{-6}$ units respectively.
  Based in the degree of stenosis, this resulted in $\sim 700$ million cells.
  When the pipe had the minimum degree of stenosis ($25\%$), there was more
  volume of the fluid resulting in approximately 1 million more lattice cells
  than the most severe case.
  Presuming the main vessel diameter, D, of $1$ unit, the spatial resolution
  $\delta x$ resulted in $124$, $156$ and $232$ cells along the diameters of
  the throat for $60\%$, $50\%$ and $25\%$ degree stenoses respectively.
  The number of cells along the main diameter of the parent pipe was $312$.
  This ensured the symmetry of the mesh.
  {\color{black}
    It may be remarked that the number of lattice cells here does not translate
    directly to the \emph{order} of polynomial or \emph{resolution in space} for
    other higher order methods like the spectral methods.
    We employed the D3Q19 stencil of the LBM in this study, which means that each
    lattice cell had 19 degrees of freedom.
    In higher order methods, the order of the polynomial allows for a higher
    accuracy within a cell, which is not the case with LBM.
    Thus these resolutions cannot be compared with those employed in similar
    studies~\citep{varghese2}.
    A detailed comparison of methods is left for future efforts.
  }
    
  Our previous works~\citep{parco, musubi, parco2016} have shown that the
  \emph{Musubi} LBM solver exhibits excellent weak and strong scaling as well
  as parallel efficiency of more than $90\%$ on supercomputers when every
  core has more than about $4000$ lattice nodes.
  Simulations were thus executed on $76\,800$ cores of the \emph{SuperMUC-NG}
  supercomputer installed at the Leibniz Supercomputing Center in Munich,
  \textsc{Germany}.
  With this setup each core (or MPI rank) had about $9\,000$ lattice nodes for
  computations thereby reducing the communication to computation ratio to an
  optimal minimum, and ensuring fully efficient utilization of compute
  resources.

  {\color{black}
    Simulations at Reynolds of $2000$ and $2100$ are only reported in this work
    due to the laminar nature of the flow at lower Reynolds numbers.
    The simulations for all the cases were first conducted for only $n=4$
    cycles as in case of laminar flow, computation of more cycles was
    redundant.
    Some of these simulations were conducted using $7200$ CPUs of the Snellius
    system, which is the national supercomputer in \textsc{The Netherlands}.
    In case of a transitional or a turbulent flow, it is always challenging to
    decide how many total cycles should be used for ensemble averaging until
    the quantities are statistically converged as the compute costs and
    physical details need to be balanced.
    Ideally, the ensemble average should attain the shape of a laminar flow,
    similar to inflow or the turbulent fluctuations should wash out from the
    averaged plot.

    Due to the lack of this information \emph{a priori}, we first simulated the
    flow with pulsation frequency $\omega_1$.
    In the cases where flow transition was identified, we restarted those
    simulations from the 4th cycle onwards and allowed the computation of
    further $n=20$ cycles resulting in a total of $n=24$ cycles.
    For higher pulsation frequencies, due to the same time step, the number of
    cycles automatically became $n=48$ for $\omega_2$ and $n=96$ for
    $\omega_3$.  
    Each cycle required $\sim 4.3$ minutes of execution time.
    An \emph{abort criteria} was set in \emph{Musubi} to stop the simulation
    upon achievement of a steady ensemble average.
    Thus, each subsequent cycle was ensemble averaged with the previously
    computed cycles and when the velocity fluctuations in ensemble averages
    vanished, the simulations were stopped automatically.
    This analysis revealed that a total of about $n=20$ cycles were enough for
    the most turbulent simulation, which further implicated the choice of
    $n=20$ cycles for all the cases.
    Based on this a total of $n=22$ cycles were computed for each simulation
    and the last 20 were used for the analysis of turbulent flow
    characteristics.
    It may be noted that 20 cycles are relatively less than what would be
    needed for statistical convergence.
    A recent study by \citet{andersson21a} reports that employment of much
    larger number of cycles is required for statistical convergence.
    The 20 cycles in this case are sufficient because of the zero mean nature
    of the flow, which is absent in a pulsatile blood flow reported in the
    study of~\citet{andersson21a}.
  }

  \subsection*{Flow characterization} \label{subsec:flowanal}
  A total of $n=22$ (where initial 2 cycles are discarded from analysis) cycles
  were computed for each Re simulation.
  Velocity over these cycles was ensemble averaged for the analysis 
  of turbulent characteristics as:
        \begin{align} 
          \label{eq:phav}
          \overline{u}(x,t) = \frac{1}{n}\sum_{k=0}^{n-1} u (x,t + kT)  
        \end{align}
  where $u(x,t)$ is the instantaneous velocity at a particular lattice site,
  \textbf{x} denotes the spatial coordinates, \textbf{t} is the time and
  \textbf{T} is the period of cycle.

  We assumed that turbulence was the only source of fluctuations in the flow
  and composed the velocity field into a mean and a fluctuating component.
  This is known as Reynolds' decomposition and is commonly used for steady
  flows.
  The oscillatory flows have a reversal component in addition to the pulsating
  nature of the flow.
  Triple decomposition~\citep{hussainre} is thus considered a more suitable
  method for the analysis of such flows:
        \begin{align} \label{eq:trip_decomp}
          u_i(x,t) = \bar{u_i}(x) + u_{i}^{\prime}(x,t) + \tilde{u_{i}}(x,\phi)
        \end{align}
  Thus, a periodic component denoted by $\tilde{u_{i}}(x,\phi)$ gets added to
  Reynolds' decomposition, which is a function of the time within the cycle.
  The phase is defined as $\phi = mod(t/T,1)$ where T is the total length of
  the cycle.
  Cycle-to-cycle variations are also intrinsically taken into account by the
  method of triple decomposition.

  The Turbulent Kinetic Energy (TKE) is derived from the fluctuating components
  of velocity in $3$ directions as:
  \begin{align} \label{eq:tke}
    k = \frac{1}{2}\Big( {u_{x}^{\prime 2} + u_{y}^{\prime 2} + u_{z}^{\prime 2}} \Big)
  \end{align}
  The frequency components present in a transitional or turbulent flow can
  be analyzed from the power spectral density (PSD) of the TKE.
  The PSD was computed in this work using the Welch's periodogram method, which
  was related with Kolmogorov's energy decay to observe the inertial and
  viscous ranges.

  \subsubsection*{Kolmogorov microscales} \label{subsec:klmgrv}
  In the previous~\citep{jain20a} as well as in related work~\citep{helgeland}
  the Kolmogorov theory has been taken as a reference to estimate the quality
  of the mesh employed for DNS.
  It may be noted that the degrees of stenosis considered in this article are
  relatively lower than those in the previous work where stenosis with a $75\%$
  area reduction was chosen.
  This implies that the amount of turbulence (if any) should be lower in the
  present work, which in turn suggests that the resolution employed in
  simulations should be enough for the accurate capture of transitional
  characteristics.
  It was however not \emph{a priori} known whether turbulence will occur in
  these cases at Reynolds of up to $2100$ or not.
  This means that simulations at even higher Reynolds number could have been
  required and thus Kolmogorov microscales were still calculated,
  
  The Kolmogorov scales are defined as spatial and temporal scales when
  viscosity dominates and the TKE is dissipated into heat~\citep{pope}.

  The Kolmogorov scales, non-dimensionalized with respect to the velocity scale
  $u_c$ and the length scale D are computed from the \emph{fluctuating}
  component of the non-dimensional strain rate defined as:
        \begin{align} \label{eq:flstr}
          s^{\prime}_{ij} = \frac{1}{2}\bigg(\frac{\partial u_{i}^{\prime}}{\partial x_j} +
          \frac{\partial u_{j}^{\prime}}{\partial x_i}\bigg) 
        \end{align}
  The Kolmogorov length, time and velocity scales are then respectively
  computed as:
      \begin{align} \label{eq:eta}
        \eta  = \bigg( \frac{1}{Re^2} \frac{1}{2 s^{\prime}_{ij}s^{\prime}_{ij}} \bigg)^{1/4}
      \end{align}
      \begin{align} \label{eq:taueta}
        \tau_{\eta} = \bigg( \frac{1}{2 s^{\prime}_{ij}s^{\prime}_{ij}}\bigg)^{1/2}     
      \end{align}
      \begin{align} \label{eq:ueta}
        u_{\eta} = \bigg( \frac{2s^{\prime}_{ij}s^{\prime}_{ij}}{Re^2} \bigg)^{1/4}
      \end{align}
  Based on these scales, the quality of the spatial and temporal resolution of
  a simulation is estimated by computing the ratio of $\delta x$ and $\delta t$
  against the corresponding Kolmogorov scales i.e.
        \begin{align} \label{eq:lp}
          l^{+} = \frac{\delta x}{\eta}
        \; \hspace{10pt}
          t^{+} = \frac{\delta t}{\tau_{\eta}}
        \end{align}

  The  $l^{+}$ and $t^{+}$ were computed using equation~\ref{eq:lp} for only
  the $60\%$ stenosis case at the highest frequency as the fluctuations were
  highest in this case. 
  The values of these scales were $1.18$ and $0.62$ respectively\footnote{A
  detailed analysis of Kolmogorov microscales has been omitted because the
  turbulence intensity in these cases is lower than the previous
  study~\citep{jain20a}. Please see the results section for more details.}.
\section{Results}       \label{sec:result}
  \begin{figure*}%
    \centering%
    \subfloat[axisymmetric]{
      \includegraphics[height=0.37\textheight]{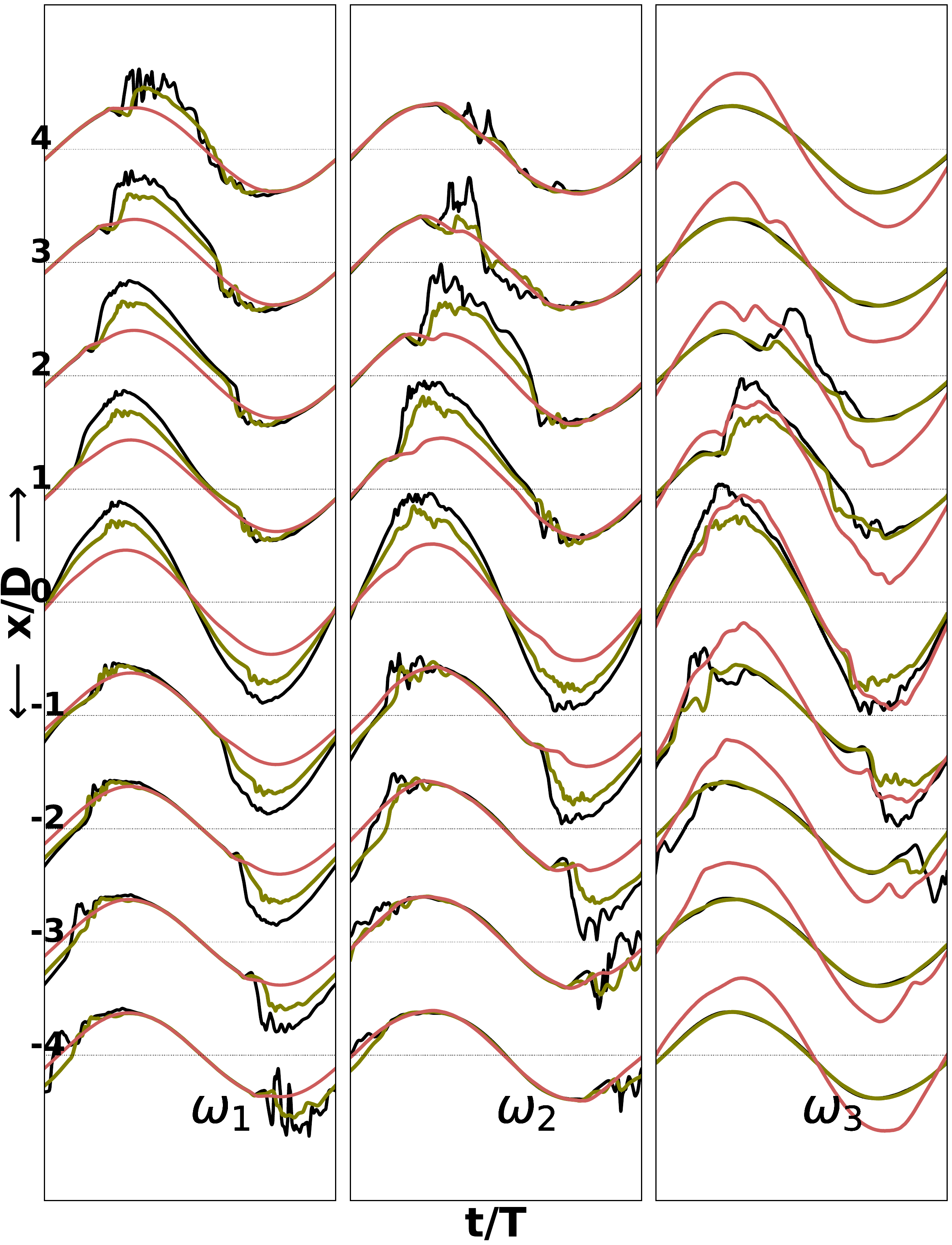}       \label{fig:axisX6}
    }
    \hspace{8pt}
    \subfloat[eccentric]{
      \includegraphics[height=0.37\textheight]{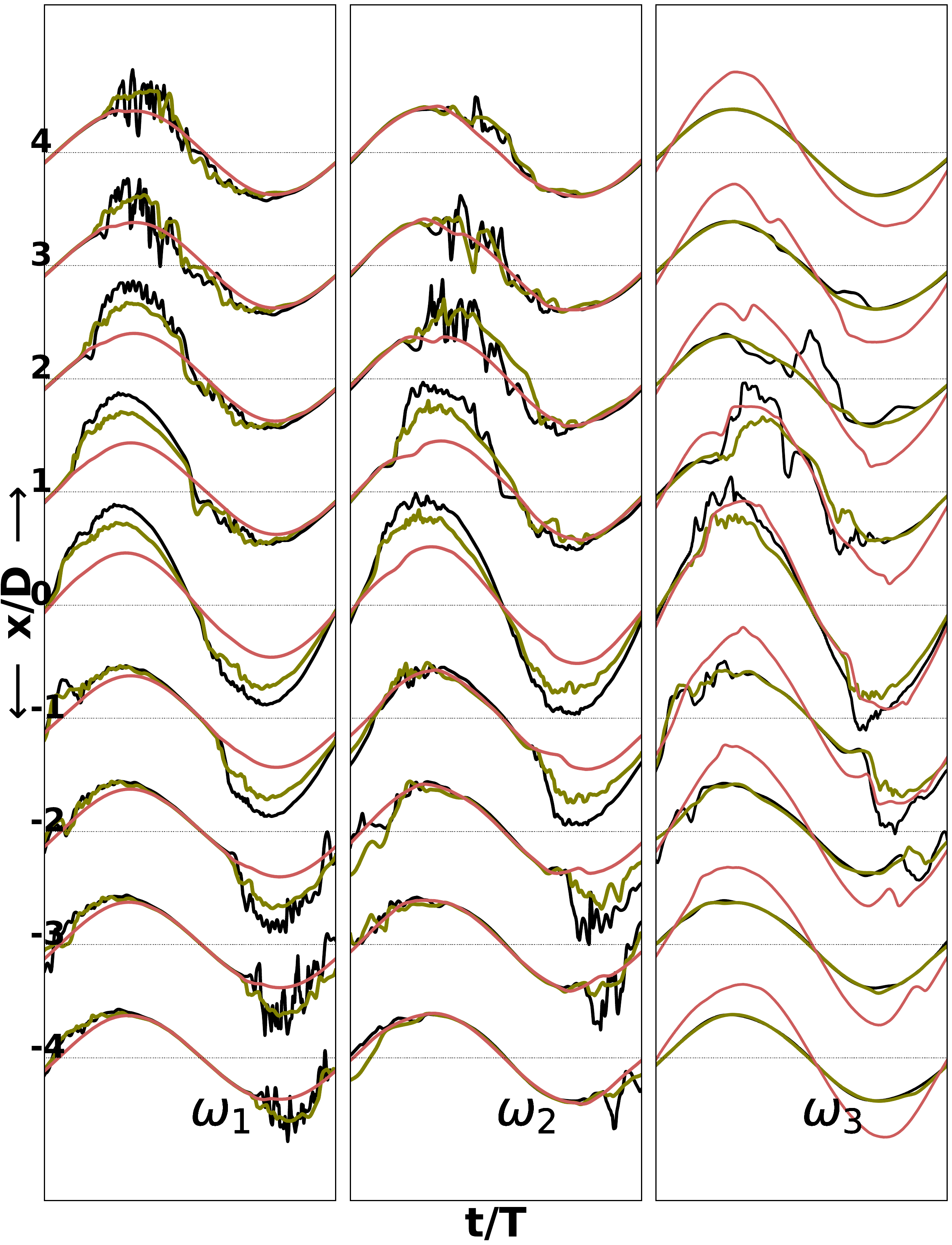}        \label{fig:eccX6} 
    }
    \caption{Normalized axial centerline instantaneous velocity during the last
    cycle for three degrees of stenoses and three oscillation frequencies in the
    axisymmetric and the eccentric cases at Re=2000.
    For oscillation frequencies, $\omega_1$ and $\omega_2$ last two and four cycles are shown.
    The black, green and red lines respectively correspond to $60\%$, $50\%$ and
    $25\%$ stenosis degrees.
    }
    \label{fig:instX6}
  \end{figure*}

  \begin{figure*}%
    \centering%
    \subfloat[axisymmetric]{
      \includegraphics[height=0.37\textheight]{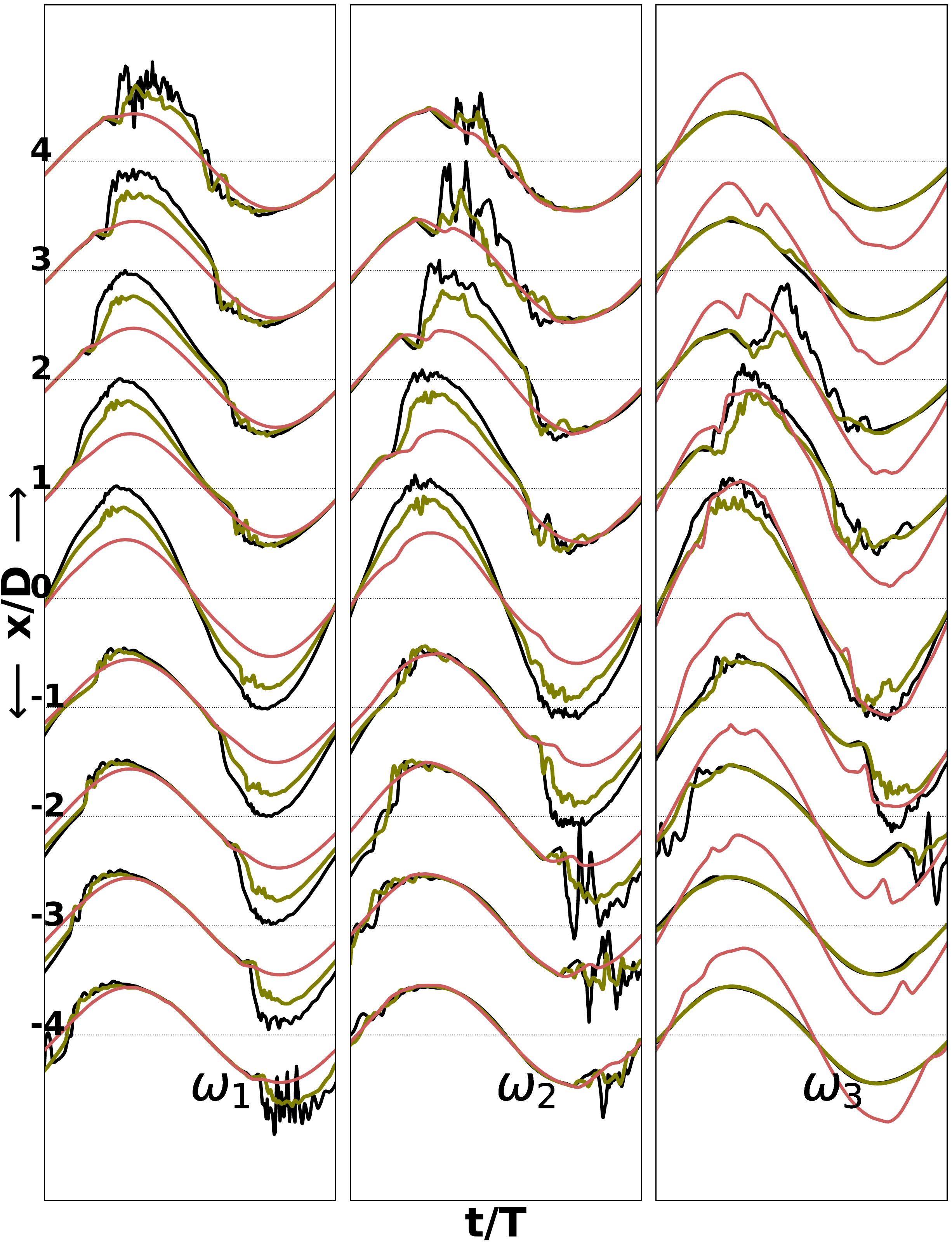}       \label{fig:axisX7}
    }
    \hspace{8pt}
    \subfloat[eccentric]{
      \includegraphics[height=0.37\textheight]{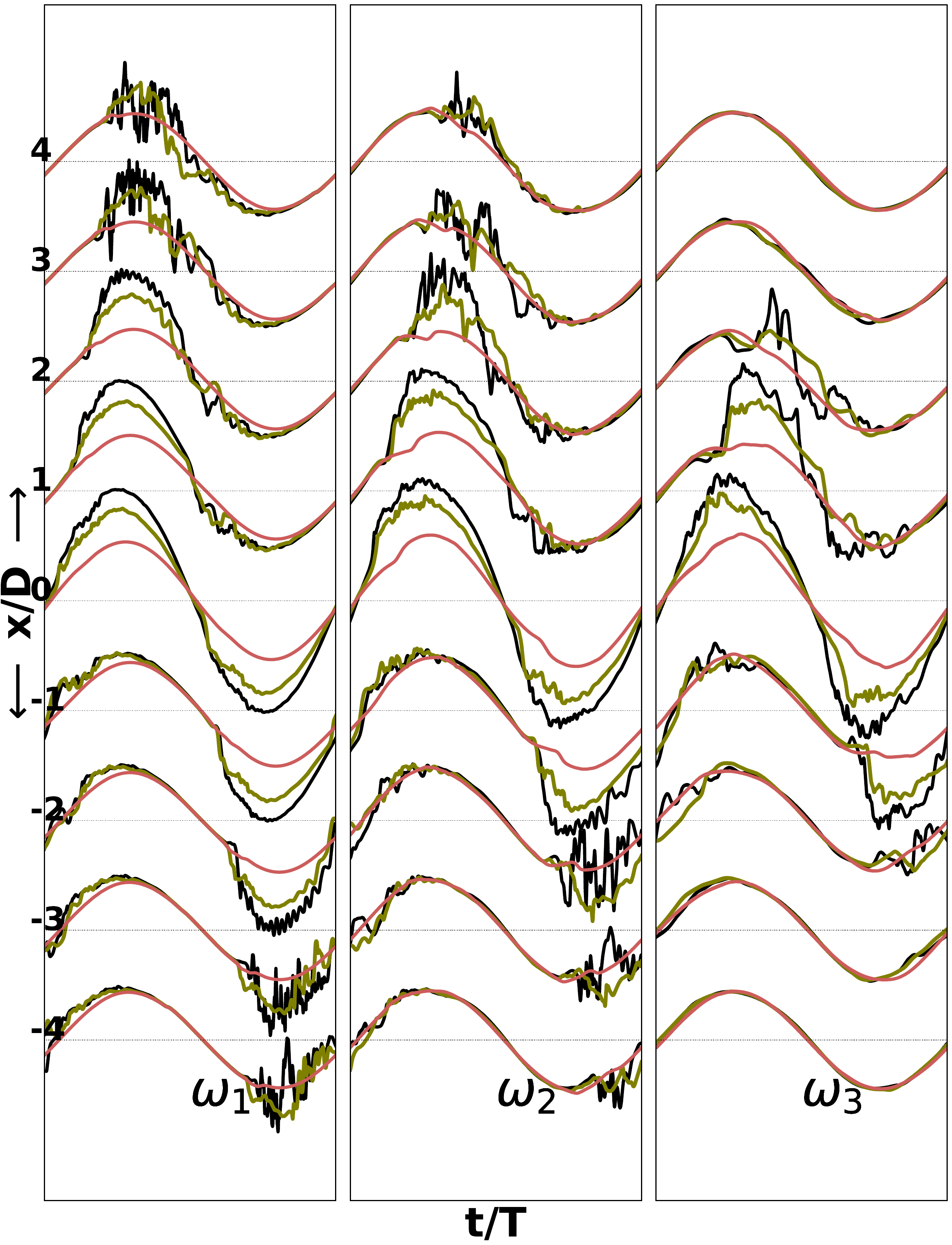}        \label{fig:eccX7} 
    }
    \caption{Normalized axial centerline instantaneous velocity during the last
    cycle for three degrees of stenoses and three oscillation frequencies in the
    axisymmetric and the eccentric cases at Re=2100.
    For oscillation frequencies, $\omega_1$ and $\omega_2$ last two and four cycles are shown.
    The black, green and red lines respectively correspond to $60\%$, $50\%$ and
    $25\%$ stenosis degrees.
    }
    \label{fig:instX7}
  \end{figure*}
  
  The flow at $Re=1800$ was laminar and it started to show valleys and peaks in
  the centerline velocities at $Re=1900$ while at $Re=2000$ these valleys and
  peaks became more pronounced and minor fluctuations started to appear mostly
  in the $60\%$ case.
  The TKE components up to this Reynolds number were depictive of a laminar flow with minor
  vortex shedding in the $60\%$ eccentric stenosis case.
  The flow only transitioned to weak turbulence at $Re=2100$ commensurate with the previous study~\citep{jain20a} of $75\%$ stenosis.
  Figures~\ref{fig:instX6} and~\ref{fig:instX7} shows traces of instantaneous centerline velocity along the streamwise
  direction for axisymmetric and eccentric cases at $Re=2000$ and $Re=2100$ respectively.

  Clearly the fluctuations at $Re=2100$ are markedly higher than those at $Re=2000$ for the $60\%$ and $50\%$ degrees of stenosis.
  The flow in both stenosis configurations is essentially laminar for the $25\%$ stenosis.
  At a higher oscillation frequency of $\omega_3$, minor distortions in the sinusoidal profile however start to appear even in the $25\%$ case.
  For the $50\%$ stenosis case the flow shows minor fluctuations during the peak flow and the deceleration phase.
  In the eccentric case these fluctuations are relatively higher at this degree of stenosis.
  The flow attains a transitional character only at a stenosis degree of $60\%$ for both the stenosis configurations.
  The fluctuations are relatively higher in the eccentric case.
  The regions downstream of the stenosis where the flow transitions and thereafter stabilizes appears to shift closer to the stenosis throat with increasing oscillation frequency.
  For the eccentric stenosis, the regions of maximum chaos are mostly confined in the locations $2D<x<4D$, $1D<x<3D$, $0D<x<2D$
  respectively for oscillation frequencies $\omega_1$, $\omega_2$ and $\omega_3$.
  In the axisymmetric case, however, the regions of jet breakdown are shifted further downstream by one diameter whereas the regions of restabilization
  are the same as in eccentric stenosis for all the studied frequencies.
  At the highest studied frequency, $\omega_3$, there are signs of fluctuations even at the stenosis throat (x=0D), which
  are a result of fast oscillation of the fluid that transfers the vortices on both sides of the stenosis rapidly.

  \begin{figure*}%
    \centering%
    \subfloat[axisymmetric]{
      \includegraphics[width=0.8\textwidth]{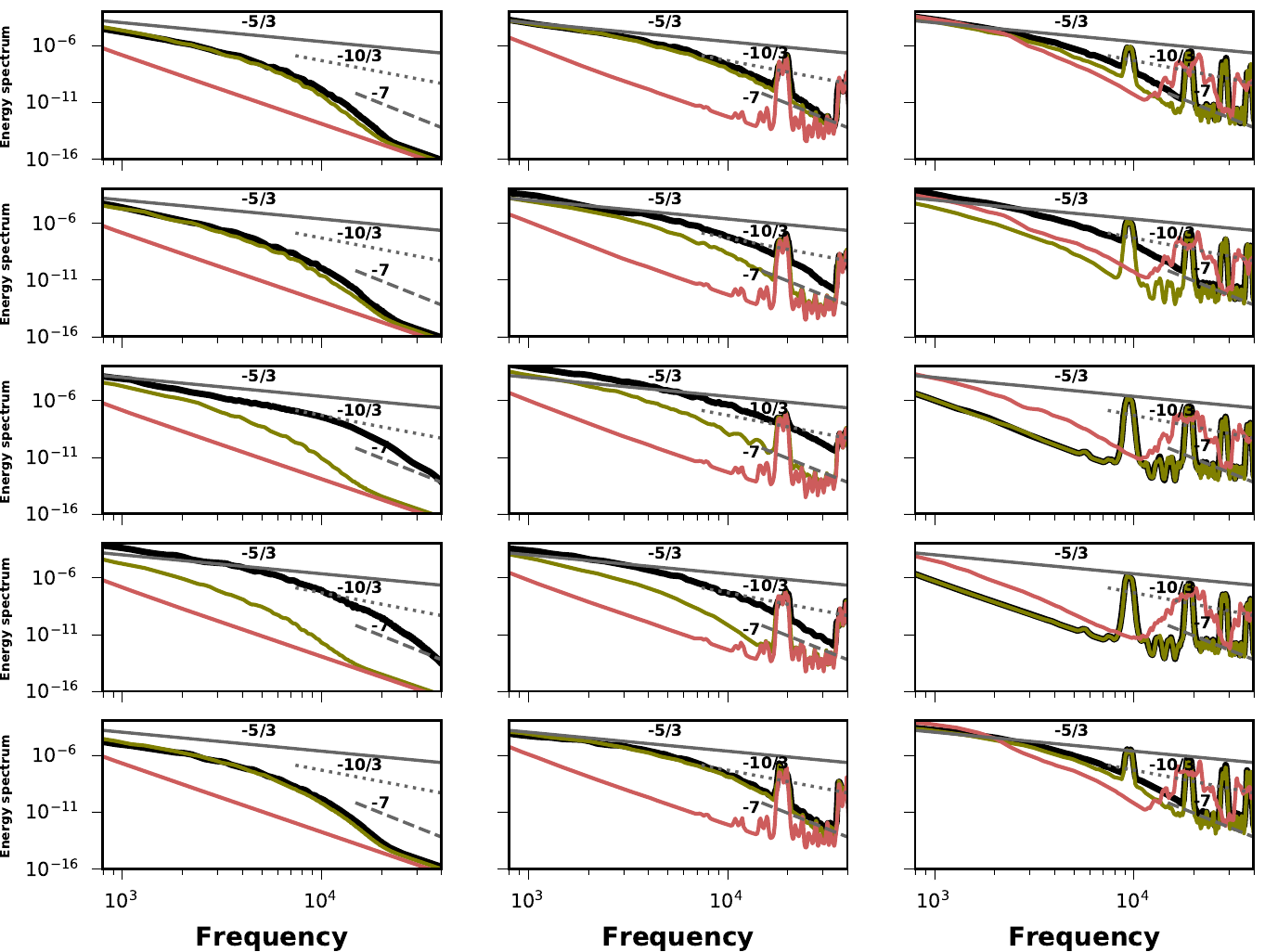}        \label{fig:axispsd}
    }
    \\
    \subfloat[eccentric]{
      \includegraphics[width=0.8\textwidth]{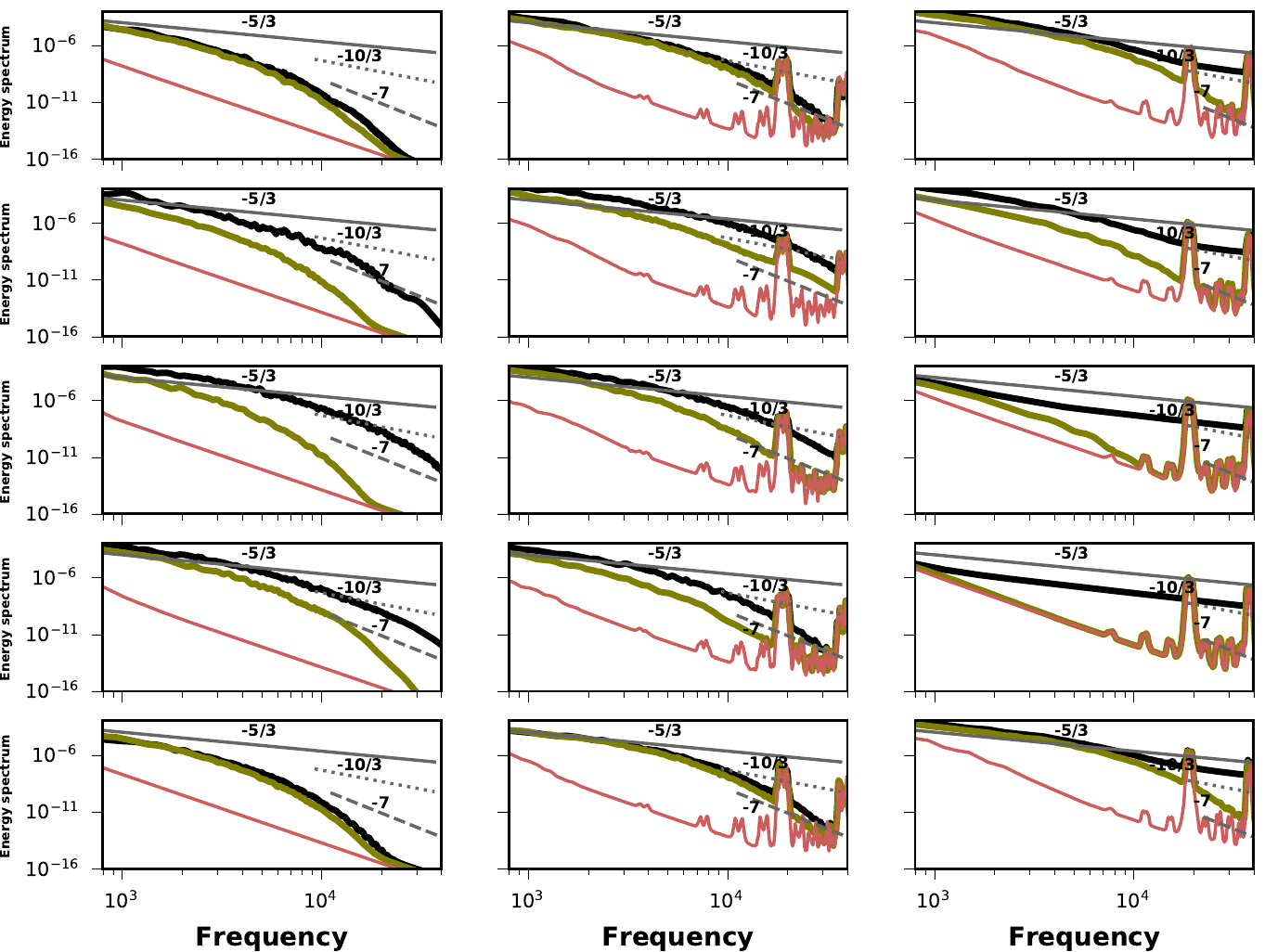}         \label{fig:eccpsd}
    }
    \caption{Power spectral density of the turbulent kinetic energy in
    axisymmetric and eccentric cases at $Re=2100$. 
    The black, green and red lines respectively correspond to $60\%$, $50\%$
    and $25\%$ stenosis degrees.  Each column shows the PSD at x=1,2,3 and 4
    diameters downstream of stenosis respectively during the blowing stage.
    Each rows from left to right shows PSD for oscillation frequencies of
    $\omega_1$, $\omega_2$ and $\omega_3$ respectively.
    The solid, dotted and dashed lines have slope of $\frac{-5}{3}$,
    $\frac{-10}{3}$ and $-7$. 
    }
    \label{fig:psd}
  \end{figure*}

  These findings are further elaborated in the power spectral density (PSD) plots of figure~\ref{fig:psd}.
  Each column from top to bottom shows the PSD of the TKE at the stenosis
  throat and 4 locations downstream of it during the blowing stage whereas the
  black, green and red lines respectively demonstrate the PSD for $60\%$,
  $50\%$ and $25\%$ degrees of stenoses.
  In the axisymmetric case for oscillation frequency $\omega_1$, the spectrum
  at $x=2D$ and $x=3D$ begins to take a shape with few frequencies in the
  inertial range depicted by the $-5/3$ decay. 
  At frequencies higher than $10^3$ the spectrum rolls of to $-10/3$ and subsequently to $-7$.
  The spectrum at $x=4D$ is similar for $60\%$ and $50\%$ degree of stenosis while the flow in $25\%$ case
  is laminar.
  Large valleys and peaks (red line) are seen in $50\%$ and $25\%$ stenoses at
  higher oscillation frequencies of $\omega_1$ and $\omega_2$
  depictive of the minor disturbances that were seen in instantaneous traces as well.
  Similar observations can be drawn from the PSD plots of the eccentric case with a clear indication of larger
  turbulent activity.
  There are larger number of frequencies in the $-5/3$ range and the roll of to $-10/3$ and $-7$ happens later in case of eccentric stenosis.

  \subsubsection*{Vortex structures}
  \begin{figure*}%
    \centering%
    \subfloat[$\asymp 60\% \sim \omega_1$]{
      \includegraphics[width=\textwidth]{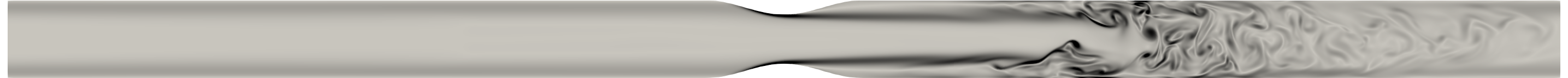} \label{fig:vax60w0}
    }\\
    \subfloat[$\asymp 60\% \sim \omega_2$]{
      \includegraphics[width=\textwidth]{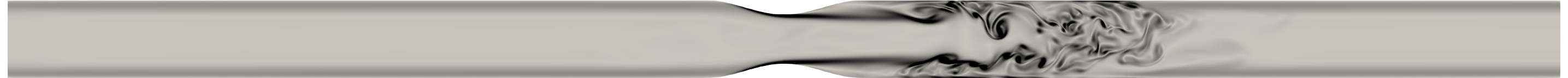} \label{fig:vax60w4}
    }\\
    \subfloat[$\asymp 60\% \sim \omega_3$]{
      \includegraphics[width=\textwidth]{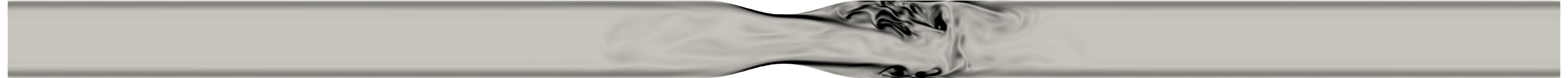} \label{fig:vax60w8}
    }\\
    \subfloat[$\asymp 50\% \sim \omega_1$]{
      \includegraphics[width=\textwidth]{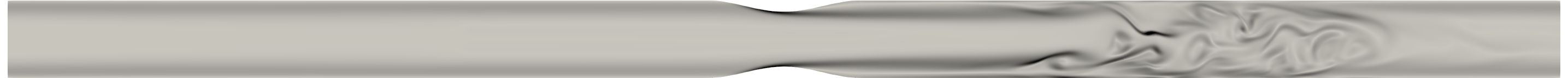} \label{fig:vax50w0}
    }\\
    \subfloat[$\asymp 50\% \sim \omega_2$]{
      \includegraphics[width=\textwidth]{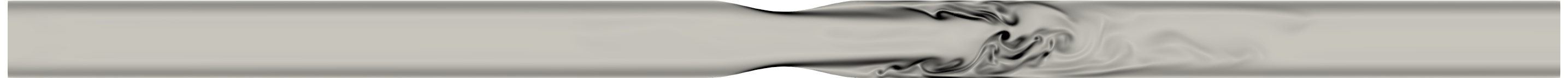} \label{fig:vax50w4}
    }\\
    \subfloat[$\asymp 50\% \sim \omega_3$]{
      \includegraphics[width=\textwidth]{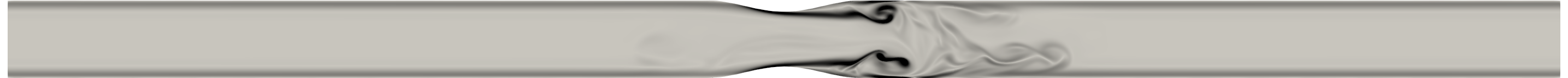} \label{fig:vax50w8}
    }\\
    \subfloat[$\asymp 25\% \sim \omega_1$]{
      \includegraphics[width=\textwidth]{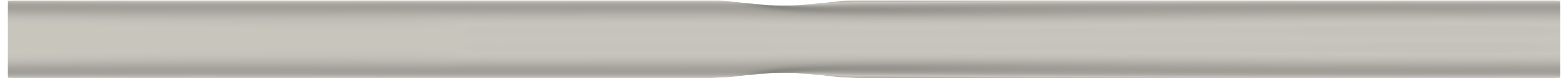} \label{fig:vax25w0}
    }\\
    \subfloat[$\asymp 25\% \sim \omega_2$]{
      \includegraphics[width=\textwidth]{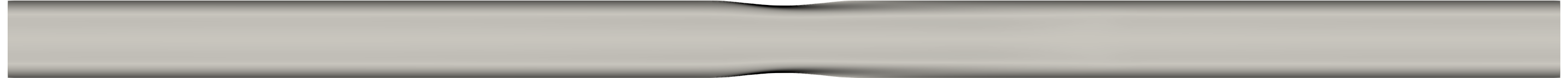} \label{fig:vax25w4}
    }\\
    \subfloat[$\asymp 25\% \sim \omega_3$]{
      \includegraphics[width=\textwidth]{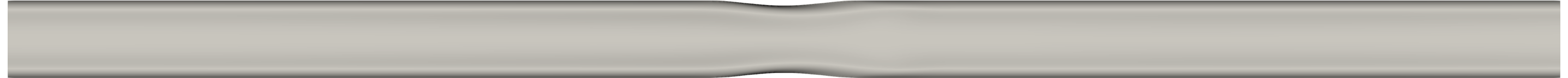} \label{fig:vax25w8}
    }\\
    \subfloat{
      \includegraphics[width=0.3\textwidth]{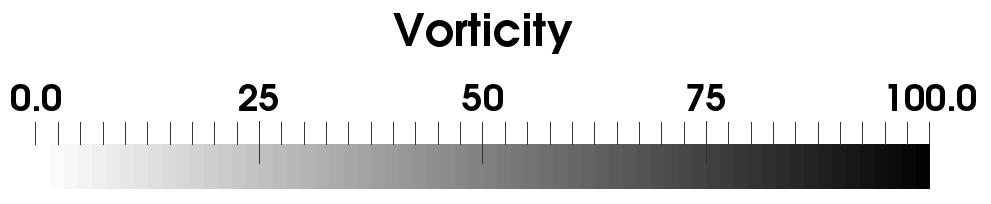}
    }
    \caption{Instantaneous vorticity magnitude at the observation point P3
    across bisecting planes in the axisymmetric stenoses of various degrees for
    $Re=2100$ and oscillation frequencies $\omega_1$, $\omega_2$ and $\omega_3$. 
    The vorticity is normalized by $u_c/D$.}
    \label{fig:axisV}
  \end{figure*}

  \begin{figure*}%
    \centering%
    \subfloat[$\asymp 60\% \sim \omega_1$]{
      \includegraphics[width=\textwidth]{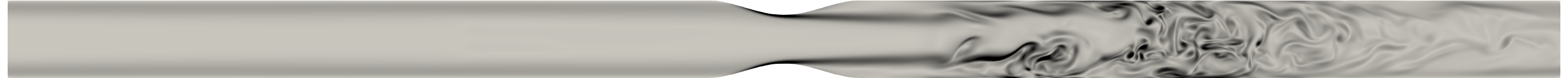} \label{fig:vec60w0}
    }\\
    \subfloat[$\asymp 60\% \sim \omega_2$]{
      \includegraphics[width=\textwidth]{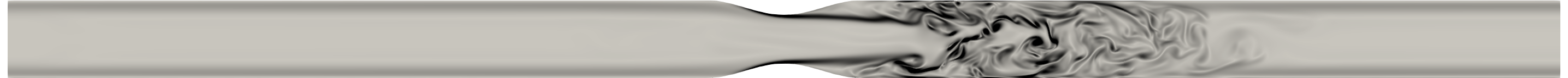} \label{fig:vec60w4}
    }\\
    \subfloat[$\asymp 60\% \sim \omega_3$]{
      \includegraphics[width=\textwidth]{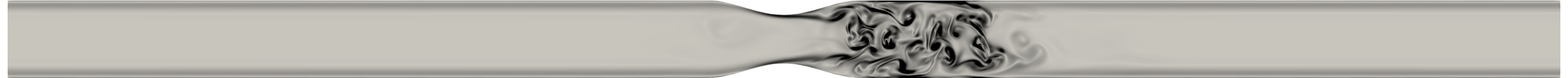} \label{fig:vec60w8}
    }\\
    \subfloat[$\asymp 50\% \sim \omega_1$]{
      \includegraphics[width=\textwidth]{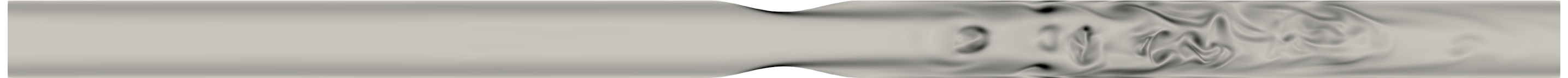} \label{fig:vec50w0}
    }\\
    \subfloat[$\asymp 50\% \sim \omega_2$]{
      \includegraphics[width=\textwidth]{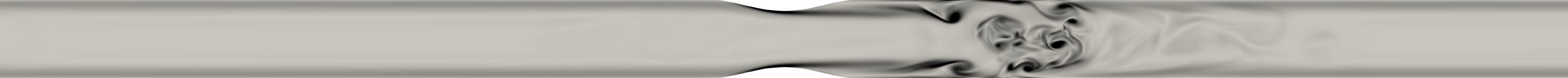} \label{fig:vec50w4}
    }\\
    \subfloat[$\asymp 50\% \sim \omega_3$]{
      \includegraphics[width=\textwidth]{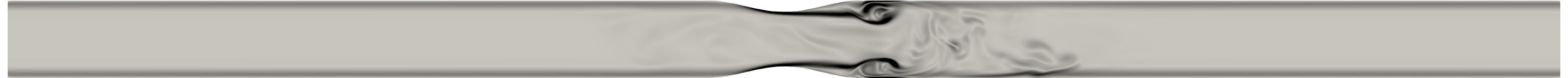} \label{fig:vec50w8}
    }\\
    \subfloat[$\asymp 25\% \sim \omega_1$]{
      \includegraphics[width=\textwidth]{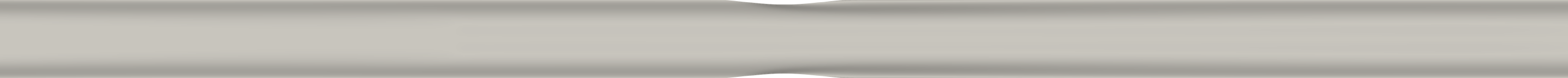} \label{fig:vec25w0}
    }\\
    \subfloat[$\asymp 25\% \sim \omega_2$]{
      \includegraphics[width=\textwidth]{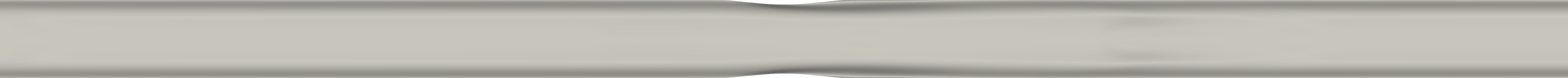} \label{fig:vec25w4}
    }\\
    \subfloat[$\asymp 25\% \sim \omega_3$]{
      \includegraphics[width=\textwidth]{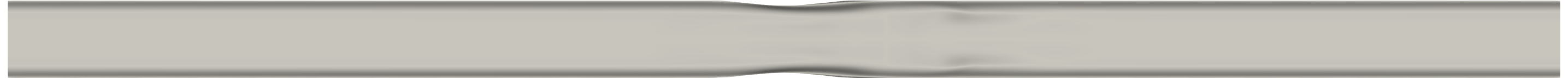} \label{fig:vec25w8}
    }\\
    \subfloat{
      \includegraphics[width=0.3\textwidth]{vortleg.png}
    }
    \caption{Instantaneous vorticity magnitude at the observation point P3
    across x-z bisecting planes in the eccentric stenoses of various degrees for
    $Re=2100$ and oscillation frequencies $\omega_1$, $\omega_2$ and $\omega_3$. 
    The vorticity is normalized by $u_c/D$.}
    \label{fig:eccV}
  \end{figure*}

  Figures~\ref{fig:axisV} and~\ref{fig:eccV} show the profiles of vorticity
  magnitude across a bisecting plane in the axisymmetric and the x-z bisecting
  plane in the eccentric stenoses cases of all variants and oscillation
  frequencies. 
  The observation point P3 (see figure~\ref{fig:sinus}) has been chosen as the flow exhibits maximum chaos at this observation point in the sinusoidally oscillating flow.
  The suction stages (points P4{\textendash}P6) have been omitted because more or less
  similar profiles are seen on the left side of the stenosis
  throat\footnote{Please refer to the animations in supplementary material for
  more details}.
  The previous observation that the regions of jet breakdown and restabilization shift closer to the stenosis
  throat with increasing oscillation frequencies are clearly seen in the vorticity plots.
  
  In the axisymmetric case the jet that emanates from the throat has enough time to travel downstream at oscillation frequency $\omega_1$
  before its breakdown and transition of flow to weak turbulence.
  At higher frequencies, the flow breaks down closer to the throat as the deceleration phase commences sooner in these cases.
  It may be specifically seen that at $60\%$ stenosis and oscillation frequency
  $\omega_3$ (figure~\ref{fig:vax60w8}) the jet is deflected towards the lower
  part downstream of the stenosis in this particular cycle.
  This is a consequence of the fact that when the flow reverses its direction, the vortices
  merge just downstream of the throat.
  The frequency ($\omega_3$) is relatively high that breaks the symmetry that is
  otherwise present at lower frequencies.
  The momentum of the flow is suddenly lost when it reverses its direction and this happens at a rapid speed
  at this frequency resulting in this loss of symmetry even in the axisymmetric case.
  This is also seen to a certain extent in $50\%$ stenosis with $\omega_3$ (figure~\ref{fig:vax50w8}) albeit at a lower extent due to the
  lower intensity of turbulence in the first place due to less constriction in the pipe.
  For the $25\%$ stenosis case this phenomenon is never observed as the flow is clearly laminar for all the oscillation frequencies. 

  The eccentricity of the stenosis causes the jet to deflect earlier than the axisymmetric case 
  resulting in a jet breakdown location closer to the stenosis throat.
  Here a higher oscillation frequency amplifies the chaos and factors that
  influence the onset of transition, in addition to eccentricity, are faster
  blowing and suction as was stated above.
  For the $50\%$ stenosis case at oscillation frequency $\omega_1$ (figure~\ref{fig:vec50w8}), the jet
  does not appear to be deflected as a result of eccentricity but an inflexion point is created
  about $2.5$ diameters downstream of stenosis where the turbulent activity occurs and extends beyond to up to about $4$ diameters.
  At higher frequencies in this degree of stenosis, interesting patterns are observed with a similar fluid dynamical behavior as in axisymmetric stenosis.

\section{Discussion}

The direct numerical simulations of oscillatory flow in varying degrees of
stenosis in axisymmetric and eccentric configurations have found a correlation between
severity of the stenosis and the intensity of turbulence in an oscillatory flow.
This influence is a well established fact but none of the previous studies have
characterized this for oscillatory flow.

  In comparison to the previous study with $75\%$ stenosis~\citep{jain20a},
  this study has contrastingly found that the increasing frequencies of
  oscillation result in earlier breakdown of the flow, a phenomenon most
  pronounced in lower degrees of stenosis.
  In a constriction as large as $75\%$ the onset of turbulence was predominated
  by the geometry rather than the complexity of the flow field.
  Upon reducing the constriction in the present study the frequencies of the
  oscillatory flow field resulted in the onset and sustainment of turbulent
  flow.
  At high frequencies of oscillation, the instant at which the downstream flow
  reversed direction had a lag in different radial directions resulting in
  merger of vortices and creation of newer ones.
  
  The axisymmetric or the eccentric nature of the geometry here also had less
  pronounced effect in comparison to the previous study, mainly because of the
  aforementioned factors.
  The most interesting distinction between eccentric and axisymmetric cases is
  seen in $25\%$ degree of stenosis and pulsation frequency $\omega_3$.
  As can be seen from figures~\ref{fig:vec25w4} and~\ref{fig:vec25w8} miniature
  patches of vortices are created downstream of the stenosis.
  These vortices that are shed are a consequence of jet deflection, which sows
  the seeds of flow transition already at this small constriction in the pipe.
  These seeds are not intense enough to cause a breakdown of the flow at this
  Reynolds number, and the flow reversal further adds to their incapability in
  bringing about flow transition.

  The intermittent nature of turbulence along the flow cycle demonstrates how
  flow reversal stabilizes the flow field {\textendash} commensurate with the
  previous study~\cite{jain20a}.
  The centerline velocities however demonstrate different characteristics
  with different oscillation frequencies.
  At larger pulsation frequencies $\omega_2$ and $\omega_3$ the turbulence has
  more a decaying nature because the deceleration phase has a reduced time with
  increasing frequencies.

  One of the most surprising observations stems from a comparison of figures
  \ref{fig:vax60w8} and \ref{fig:vec60w8}.
  While the former figure is for axisymmetric case and the latter for eccentric
  case, the shape of the flow jet seems to suggest as if these figures have
  been interchanged.
  The time lapse between events like flow jet breakdown, reversal and stabilization is 
  extremely small in these cases due to the high frequency of the sinusoidal waveform.
  Due to that the vortices that were created during the deceleration phase are immediately
  disturbed by the flow reversal that tries to stabilize the flow field and push the flow in the
  other direction.
  A fast merger and annihilation of vortices breaks the shape of the post stenotic jet
  giving rise to this behavior.
  If we take a closer look at figure~\ref{fig:vec60w8} we can observe that the
  jet is actually deflected upwards due to the eccentricity\footnote{Such
  observations are more clear in the animations provided as supplementary
  information}.

  From the PSD plots (figure~\ref{fig:psd}) it may be seen that whatever
  turbulent nature of the flow there is, it is less intense than that for
  $75\%$ stenosis.
  At higher degrees of stenoses the PSD attains a $\frac{-5}{3}$ decay albeit
  for a very small range of frequencies. 
  A roll of to $\frac{-10}{3}$ and $-7$ happens much sooner demonstrating the
  decaying nature of turbulence.

  The degree of constriction, in almost all pathologies indicates the severity of the situation.
  \citet{bluestein95a} focused their study with background of heart valves and for pulsating flow
  their general observation, that a higher constriction results in larger turbulence, was consistent
  with the results of the present study.
  It may be queried how relevant a zero mean oscillatory flow is in physiology.
  \citet{holenstein88a} has demonstrated the relevance of reverse flow in infrarenal vessels as well
  albeit the flow there only has a reversed component and the mean of the flow is not zero.
  This study will thus have a main relevance in CSF and respiratory flow research that are purely oscillatory. 

  {\color{black}
  This study has several limitations, most of which are similar to the previous
  study~\citep{jain20a}.
  In the context of degrees of stenosis it may be mentioned that despite the
  study of three different pulsation frequencies, the waveform has always been
  sinusoidal.
  In both CSF and respiratory flows, the waveforms are very complex due to a
  number of physiologic factors.
  Such waveforms might result in alternate turbulence mechanisms.
  Also, despite eccentricity the stenosis shape is still symmetric, which is
  unlikely to happen in an anatomic case.
  This study may thus be inferred as one where complexity of the anatomy has
  been incorporated in the form of major factors that are present, while
  maintaining the constitutive simplicity so that these results can act as a
  benchmark.
  }

\subsection*{Acknowledgements}
    Compute resources on \emph{SuperMUC} and \emph{SuperMUC-NG} were provided
    by the Leibniz Supercomputing Center (LRZ), Munich, \textsc{Germany}.
    The author is grateful for the able support of colleagues at LRZ.
    Some simulations were performed on the Dutch national supercomputer
    Snellius.
    Compute resources on Snellius were provided by the SURFSara through NWO
    grant 2019/NWO/00768083.

\subsection*{Conflict of Interest Statement}
  The author declares that he has no conflict of interest.
  No funding was received for this specific work.

\bibliographystyle{apalike2}       
\bibliography{references}%

\end{document}